# Evidence for hard and soft substructures in thermoelectric SnSe

S. R. Popuri,[a] M. Pollet,[b,c] R. Decourt,[b,c] M. L. Viciu[d] and J. W. G. Bos[a*]

[a] Institute of Chemical Sciences and Centre for Advanced Energy Storage and Recovery, School of Engineering and Physical Sciences, Heriot-Watt University, Edinburgh, EH14 4AS, United Kingdom.
[b] CNRS, ICMCB, UPR 9048, Pessac F-33600, France
[c] University of Bordeaux, UPR 9048, Pessac F-33600, France
[d] Department of Chemistry and Applied Biosciences, ETH Zürich, Vladimir-Prelog-Weg 1, Zürich CH 8093, Switzerland

j.w.g.bos@hw.ac.uk; michael.pollet@icmcb.cnrs.fr

SnSe is a topical thermoelectric material with a low thermal conductivity which is linked to its unique crystal structure. We use low-temperature heat capacity measurements to demonstrate the presence of two characteristic vibrational energy scales in SnSe with Debye temperatures $\theta_{D1}$ = 345(9) K and $\theta_{D2}$ = 154(2) K. These hard and soft substructures are quantitatively linked to the strong and weak Sn-Se bonds in the crystal structure. The heat capacity model predicts the temperature evolution of the unit cell volume, confirming that this two-substructure model captures the basic thermal properties. Comparison with phonon calculations reveals that the soft substructure is associated with the low energy phonon modes that are responsible for the thermal transport. This suggests that searching for materials containing highly divergent bond distances should be a fruitful route for discovering low thermal conductivity materials.

Thermoelectric generation is a promising technology that enables waste heat to be converted in useful electricity.[1] There are many possible applications but widespread use has been limited by the lack of cost effective efficient thermoelectric materials. The underpinning problem is the need to optimise competing materials parameters. This is commonly expressed by the thermoelectric figure of merit, $zT = (S^2/\rho\kappa)T$, where S and $\rho$ are the Seebeck coefficient and electrical resistivity, $\kappa$ is the sum of the lattice and electronic thermal conductivities, and T is the absolute temperature. Many of the current best thermoelectric materials are based on low lattice thermal conductivities. For example, PbTe-based nanocomposites routinely achieve $\kappa < 1$ W m$^{-1}$ K$^{-1}$ and ZT values between 1.5 and 2 are now common.[2-4] A common factor in these materials with ultralow thermal conductivities is the presence of chemically inert lone-pairs.[5,6] Recently, outstanding thermoelectric performance (ZT = 2.5) was discovered in SnSe single crystals without any carrier doping to optimise the power factor or nanostructuring to minimize the thermal conductivity.[7] The large ZT values are primarily based on an ultralow lattice thermal conductivity $\kappa_{lat} = 0.2$-$0.3$ W m$^{-1}$ K$^{-1}$ at 800-900 K. This was attributed to the highly anharmonic bonding, leading to very strong phonon-phonon scattering, and the anisotropic crystal structure.[7-12] The crystal structure can be understood as a highly distorted rock salt derivative.[13,14] At high-temperatures, adjacent two-atom thick rock salt layers are translated by a/2 so that Sn is coordinated by 5 Se atoms in the rocksalt layer and by 2 Se atoms in an adjacent layer. Upon cooling below 800 K, the structure undergoes a displacive phase transition, which leads to a strongly distorted Sn coordination polyhedron with three short and two long bonds within the rocksalt layer, and a single long bond connecting to an adjacent rocksalt layer (Fig. 1). There is recent controversy regarding the intrinsic thermal conductivity in SnSe,[15,16] with a new single crystal study suggesting that $\kappa_{lat} \approx 2$ W m$^{-1}$ K$^{-1}$ within the tightly bound rocksalt layers and ~1 W m$^{-1}$ K$^{-1}$ in the perpendicular direction (300 K values).[17] In polycrystalline samples a range of $\kappa_{lat}$ values have been reported,[18-22] with some suggestion that disorder is able to reduce $\kappa_{lat}$.[21,23] The measured $\kappa_{lat}$ values have also been reported to be sensitive to oxidation.[24] However, the promise of SnSe as an outstanding thermoelectric material is not in doubt as Na-doped single crystals show greatly improved thermoelectric power factors near room temperature and compete with the best known Bi$_2$Te$_3$ based alloys in terms of their efficiency.[25,26]

In this Letter, we report an investigation into the low-temperature heat capacity of SnSe and combine this with crystal structure data to yield important insights into the thermal behaviour of SnSe. The heat capacity of SnSe was measured using a Quantum Design Physical Property Measurement System, and matched well with high-temperature data reported previously.[23] The SnSe sample used was prepared using solid state reactions followed by hot pressing and is highly pure with < 1 wt% $SnO_2$ or other impurities. The detailed synthesis protocol and thermoelectric properties are described in Ref. 23. Neutron powder diffraction data were collected using the HRPD instrument at ISIS, Rutherford Appleton Laboratory, UK. A full structural study will be published elsewhere[27] and only the obtained cell volume and thermal displacement parameter are reported here.

The temperature dependence of the thermal displacement parameters gives a first insight into the vibrational properties of Sn and Se. Attempts to use anisotropic displacement parameters revealed negligible anisotropy and for this reason the isotropic values are used here. The temperature dependence of the isotropic displacement parameters ($U_{iso}$) for Sn and Se are shown in Fig. 1b. The $U_{iso}$ values are large, in particular for Sn, which is typical for lone-pair containing rocksalt-based thermoelectrics.[28] The $U_{iso}$'s tend toward zero at low temperatures, demonstrating that the structure does not contain significant static disorder.

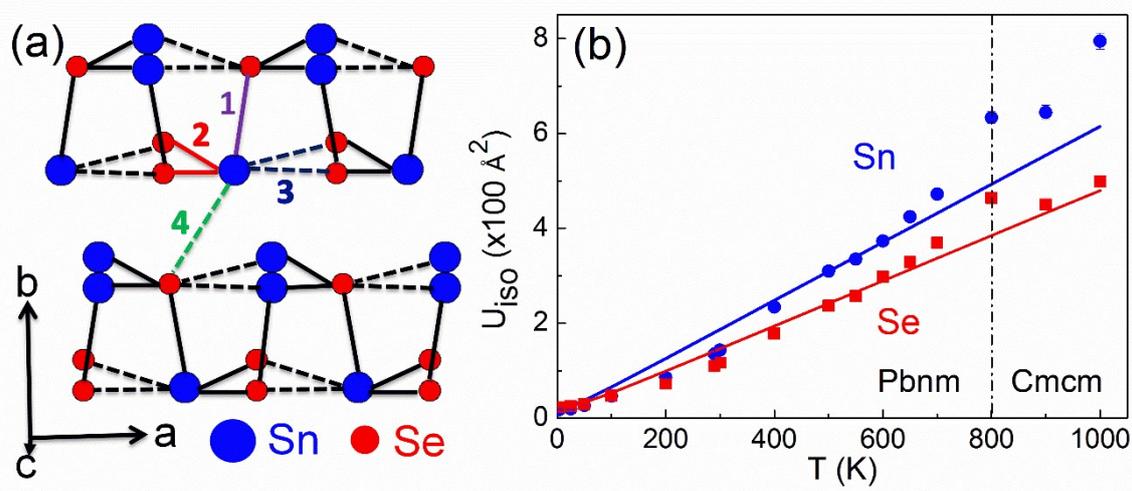

Fig. 1. (a) Schematic of the low-temperature crystal structure of SnSe (Pbnm setting). Important bond distances are indicated. The 300 K values for the bond distances are as follows: $d_1$ = 2.75 Å; $d_2$ = 2.79 Å; $d_3$ = 3.34 Å and $d_4$ = 3.47 Å. (b) Fit to the temperature dependence of the isotropic thermal displacement parameters ($U_{iso}$) for Sn and Se using a Debye model.

In order to obtain an estimate of the characteristic energy scales of the atomic motions the $U_{iso}$'s for Sn and Se were fitted below 800 K using:[29, 30]

$$U_{iso} = \frac{3\hbar^2}{mk_B\theta_D}\left[\frac{1}{4} + \left(\frac{T}{\theta_D}\right)^2 \int_0^{\theta_D/T} \frac{x}{e^x-1}dx\right] + \sigma^2 \quad (1)$$

Here, $\hbar$ is the reduced Planck constant, $k_B$ is Boltzmann's constant, $\theta_D$ is the Debye temperature and $\sigma^2$ is the displacement correlation function. In both cases $\sigma^2 = 0$, confirming the absence of significant static structural disorder. The fitted $\theta_D$ values are 140(2) K for Sn and 195(3) K for Se, and the fits are shown as solid lines in Fig. 1. The 40% larger value for Se is in line with expected trends based on the atomic mass and bond strength. The $U_{iso}(T)$ show considerable non-linearity in the Pnma phase. This is particularly evident for Sn and reflects the highly anharmonic bonding in this material.

The temperature dependence of the heat capacity ($C_p$) is given in Fig. 2a, and is characterised by a linear increase beyond the Dulong-Petit value of 3R/atom, consistent with the high-temperature data in the literature.[7, 9, 19, 23] A plot of $C_p/T$ versus $T^2$ reveals that there is no electronic contribution to the heat capacity at low-temperature, which is in keeping with the semiconducting nature of SnSe (Fig. 2b). A plot of $C_p/T^3$ versus T reveals an additional low-temperature contribution, which could arise from the contribution of either an Einstein mode or a Schottky anomaly to the specific heat (Fig. 2c).

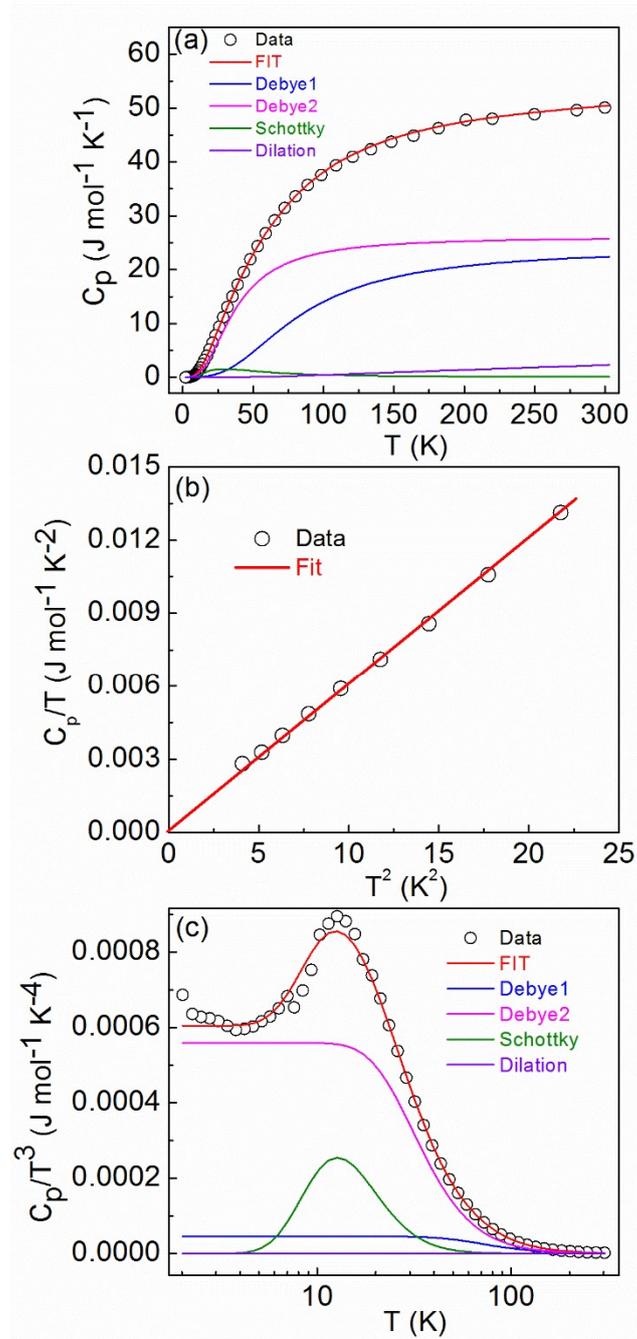

Fig. 2: (a) Fit to the temperature dependence of the heat capacity ($C_p$) of SnSe using two Debye terms (blue and pink line), Schottky mode (green line) and lattice dilation (purple line). (b) $C_p/T$ versus $T^2$ illustrating the absence of a significant electronic contribution to $C_p$. (c) $C_p/T^3$ versus T, highlighting the Schottky contribution at low temperatures.

Several models using combinations of Debye and Einstein (or Schottky) modes were tried. The most satisfactory fit was obtained using two Debye terms for the acoustic phonon bath, a low-temperature Schottky contribution to model the peak visible on the $C_p/T^3$ versus T curve and a lattice dilation term[31] to account for the linear increase at high temperature:

$$C_p = \sum_{i=1}^{2}\left[9n_i R \left(\frac{T}{\theta_{D_i}}\right)^3 \int_0^{\theta_{D_i}/T} \frac{x^4 e^x}{(e^x-1)^2} dx\right] + R \frac{g\left(\frac{\Delta}{T}\right)^2 e^{-\Delta/T}}{(1+ge^{-\Delta/T})^2} + 9Bv\alpha^2 T \qquad (2)$$

Here, $n_i$ is the number of oscillators for each Debye term, R is the gas constant, $\theta_{Di}$ are Debye temperatures, $\Delta$ is the energy gap for a two-level Schottky system and g is the ratio of the degeneracies of the lower level to the upper level, B = 31 GPa is the isothermal bulk modulus,[32] $v$ is the volume per atom and $\alpha$ is the thermal expansion coefficient. The thermal expansion coefficient was derived from the temperature evolution of the cell volume as $\alpha = (1/V)(dV/dT)_P$. The cell volume was fitted simultaneously (Fig. 3) using the following expression, which was adapted from Hayward et al.[33] to include two Debye terms:

$$V = V_0 + a\sum_{i=1}^{2}\left[\int_0^T 9n_i R \left(\frac{T}{\theta_{D_i}}\right)^3 \int_0^{\theta_{D_i}/T} \frac{x^4 e^x}{(e^x-1)^2} dx\right] \qquad (3)$$

The fitted values are $n_1$ = 0.96(4), $n_2$ = 1.04(4), $\theta_{D1}$ = 345(9) K, $\theta_{D2}$ = 154(2) K, g = 0.38(2), $\Delta$ = 64(1) K and $V_0$ = 210.32 Å$^3$, a = 1.8 × 10$^{-4}$. The model for $C_p$ takes into account all important features of the data including the low temperature peak and the linear increase at higher temperature (Fig. 2). The V(T) data are fitted well below 600 K, while the experimental volume expands more rapidly at elevated temperatures (Fig. 3). The thermal expansion $\alpha$(T) has a broad maximum near 550-600 K (inset to Fig. 3), which is in good agreement with the onset of the structural phase transition.[13]

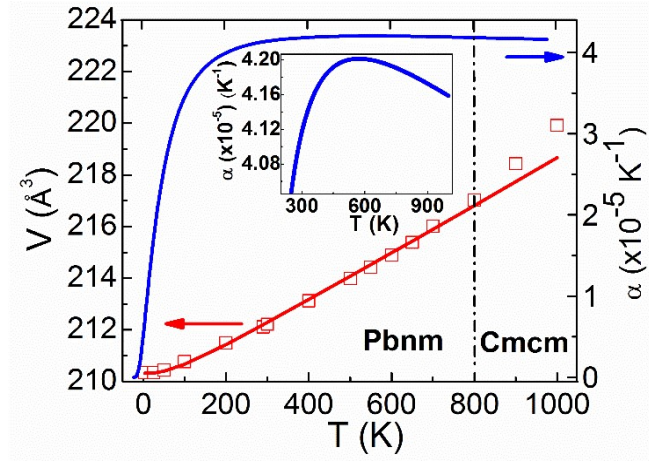

Fig. 3. Fit to the temperature dependence of the unit cell volume (V) for SnSe using the model derived from the heat capacity data. The volume thermal expansion ($\alpha$) is obtained from the fit to the volume data and shows a broad maximum near 550-600 K (see inset).

The $C_p$ fit reveals the presence of two Debye oscillators with different energy scales but almost equal weighting. It is therefore tempting to link these modes with Sn and Se atomic motions. $\theta_{D2}$ = 154(2) K matches well with the value for Sn obtained from $U_{iso}$ (140(2) K) but $\theta_{D1}$ = 345(9) K is about 75% larger than the value obtained for Se (195(3) K). This suggests that the $\theta_{Di}$ might be linked to different structural fragment and this is discussed below. The $n_i$-weighted inverse cubic average of the obtained $\theta_{Di}$ is 186(3) K, which is almost identical to $\theta_D$ = 189(2) K obtained from the low-temperature $C_p$ data using $C_p/T = \gamma + \beta T^2$ [$\beta=12\pi^4/(5\theta_D^3)$],[34] providing further validation of our model. The origin of the low-temperature Schottky contribution needs further investigation that is outside the scope of this manuscript. The availability of detailed structural and heat capacity data affords an estimate of the acoustic mode Gruneisen parameter using $\gamma = 3\alpha BV/C_v$.[35] Here, V is the molar volume $C_v$ is the constant volume heat capacity and the other terms have been defined above. The bulk modulus of SnSe is smaller than e.g. 45 and 40 GPa for PbSe and PbTe,[36] but $\alpha(T)$ is significantly larger. This yields $\gamma$ = 2.5 at 300 K, which is in good agreement with calculated values between 2-4 depending on the crystal direction.[7]

Modelling of low-temperature heat capacity data provides important insight into the link between structure and lattice dynamics. A model using two Debye oscillators of equal abundance was found to give the best fit to the data. This demonstrates that there are two important vibrational energy scales, corresponding to conceptual hard and soft substructures with an equal weighting. As discussed above, $\theta_{D1}$ = 345(9) K and $\theta_{D2}$ = 154(2) K do not directly map onto the values obtained for Sn and Se from $U_{iso}(T)$. This suggests that different structural fragments are responsible for the different vibrational energy scales observed here. The Pbnm low-temperature structure has three strong and three weak bonds with bond distances of ~2.8 Å and ~3.4 Å at 300 K. The Debye temperature (highest phonon frequency) $\theta_D \propto \sqrt{nk/m}$, where n is the number density, k is the bond strength and m is the reduced mass of the oscillator.[34] From the $C_p$ fitting, the number densities of both oscillators are equal and assuming a similar reduced mass, the ratio of the fitted $\theta_{Di}$ values is proportional to the square root of the bond strengths. We can approximate the bond strength using bond valence sums (BVS),[37] which are directly calculated from the Sn-Se bond distances. The BVS ratio is ~ 5, while the ratio of $\theta_{D1}/\theta_{D2} \approx \sqrt{5}$, signalling an almost perfect agreement between bond strength and the $\theta_{Di}$ values. The data therefore suggest that the harder substructure ($\theta_{D1}$ = 345(9) K) is linked to the short bonds, while the softer substructure ($\theta_{D2}$ = 154(2) K) is linked to the weaker bonds within and between the rocksalt layers (Fig 1a).

Computational and inelastic neutron scattering phonon studies show two discrete regions in the phonon density of states (PDOS).[8-10] A lower band spanning 0-13 meV containing 3 acoustic and 9 optic modes and a higher energy band from 13-25 meV with the remaining 12 optic modes. The upper energies for these two bands correspond very closely to highest phonon frequency for the fitted Debye modes ($k_B\theta_{D1}$ = 29.7(8) meV and $k_B\theta_{D2}$ = 13.2(2) meV). The equal number of phonon modes in the two bands in the PDOS are in agreement with the equal weighting of the Debye oscillators. It therefore is reasonable to conclude that two Debye modes correspond to the two bands in the PDOS. The phonon calculations reveal that the modes in the lower band are more strongly associated by Sn displacements, with a pronounced peak associated with motions perpendicular to the rocksalt layers.[9] The higher energy band in the PDOS is more strongly associated with Se displacements. This is consistent with our link to the weak and strong bonds in the crystal structure, where the weaker bonds allow for low-energy anharmonic Sn displacements, while Se is less strongly displaced.

Finally, we note that the presence of two lattice energy scales is similar to the skutterudite and clathrate Phonon Glass Electron Crystal (PGEC)[38] materials.[29, 39-43] These rattling systems show several characteristic lattice energy scales, typically one high corresponding to the framework and one or several low ones corresponding to weakly bound rattling cations, which are usually described using Einstein modes with $\theta_E$ < 100 K (*i.e.* $\theta_D$ = 125 K using $\theta_E/\theta_D \approx \sqrt[3]{\pi/6}$). We have demonstrated that SnSe has two lattice energy scales. The lower energy vibrational scale ($\theta_{D2}$ = 154(2) K) is of somewhat higher energy than that of a typical rattler cation, whereas $\theta_{D1}$ = 345(9) K is in the range of what is expected for a framework. The slightly higher $\theta_{D2}$ is in keeping with the different crystallography of these materials, rattlers occupy large void spaces, whereas Sn is part of the rigid framework and is only loosely bound in certain crystallographic directions. The similarity in terms of the lattice energy scales suggests that highly divergent bond distances, associated with unbalanced bond strengths and low local symmetry (e.g. the 3+3 coordination of Sn, Fig. 1a), could be another route to PGEC behavior.

To conclude, heat capacity measurements have been used to reveal that there are two characteristic vibrational energy scales in SnSe corresponding to hard and soft substructures. These distinct substructures arise because of the strong bond divergence in SnSe. Comparison to phonon calculations reveals that the soft substructure is largely responsible for the thermal transport, which is consistent with the strong Umklapp scattering and low thermal conductivities observed for SnSe. This simple link between structure and thermal properties may help with predicting new thermoelectric and thermal barrier materials. Exploring structures with widely diverging bond distances should be a fruitful route to discovering low thermal conductivity materials.

# ACKNOWLEDGEMENTS

SRP and JWGB acknowledge the EPSRC (EP/N01717X/1), Leverhulme Trust (RPG-2012-576) and the STFC for provision of beam time at ISIS. Ingo Loa is acknowledged for scientific discussions. Raw data on which this publication is based can be accessed via the Heriot-Watt University Data Repository.